\documentclass[twocolumn,preprintnumbers,superscriptaddress,aps,prl,nofootinbib,longbibliography,amsmath,amssymb]{revtex4-1}

\usepackage{graphicx}
\usepackage{dcolumn}
\usepackage{bm}
\usepackage{color}
\usepackage{ulem}
\usepackage{gensymb}
\usepackage{braket}
\usepackage{amsmath}
\usepackage[percent]{overpic}
\usepackage{lineno}

\begin{document}



\title{Anisotropic thermal conductivity oscillations in relation to the putative Kitaev spin liquid phase of $\alpha$-RuCl$_3$}

\author{Heda Zhang}
\email{zhangh3@ornl.gov}
\affiliation{Materials Science and Technology Division, Oak Ridge National Laboratory, Oak Ridge, Tennessee 37831, USA}


\author{Hu Miao}
\affiliation{Materials Science and Technology Division, Oak Ridge National Laboratory, Oak Ridge, Tennessee 37831, USA}

\author{Thomas Z Ward}
\affiliation{Materials Science and Technology Division, Oak Ridge National Laboratory, Oak Ridge, Tennessee 37831, USA}


\author{David G Mandrus}
\affiliation{Materials Science and Technology Division, Oak Ridge National Laboratory, Oak Ridge, Tennessee 37831, USA}

\author{Stephen E Nagler}
\affiliation{Neutron Science Division, Oak Ridge National Laboratory, Oak Ridge, Tennessee 37831, USA}

\author{Michael A McGuire}
\affiliation{Materials Science and Technology Division, Oak Ridge National Laboratory, Oak Ridge, Tennessee 37831, USA}

\author{Jiaqiang Yan}
\affiliation{Materials Science and Technology Division, Oak Ridge National Laboratory, Oak Ridge, Tennessee 37831, USA}

\date{\today}

\begin{abstract}
In the presence of external magnetic field, the Kitaev model could either hosts gapped topological anyon or gapless Majorana fermions. In $\alpha$-RuCl$_3$, the gapped and gapless cases are only separated by a thirty-degree rotation of the in-plane magnetic field vector. The presence/absence of the spectral gap is key for understanding the thermal transport behavior in $\alpha$-RuCl$_3$. Here, we study the anisotropy of the oscillatory features of thermal conductivity in $\alpha$-RuCl$_3$. We examine the oscillatory features of thermal conductivities (k//a, k//b) with fixed external fields and found distinct behavior for the gapped (B//a) and gapless (B//b) scenarios. Furthermore, we track the evolution of thermal resistivity ($\lambda_{a}$) and its oscillatory features with the rotation of in-plane magnetic fields from B//b to B//a. The thermal resistivity $\lambda (B,\theta)$ display distinct rotational symmetries before and after the emergence of the field induced Kitaev spin liquid phase. These experiment data suggest that oscillatory features of thermal conductivity in $\alpha$-RuCl$_3$ are closely linked to the putative Kitaev spin liquid phase and its excitations. 
\end{abstract}

\maketitle


The Kitaev model departs from the Landau paradigm of symmetry breaking and give new conservation rules by topological order (in contrast to conservation by symmetries, i.e., Noether's theorem). In this way, one can view the formation of topological order as the opposite process of symmetry breaking \cite{kitaev}. This conservation leads to a more practical perspective on the Kitaev model: that the non-Abelian anyon described within could provide a platform for (fault-tolerant) topological quantum computation \cite{kitaev, kitaevqc}. $\alpha$-RuCl$_3$ extend the abstract concept of the Kitaev model to a tangible physical system. It is a candidate material for realizing the Kitaev model, with a zigzag order antiferromagnet ground state \cite{plum,banj,rucl31,rucl32,rucl33}. The magnetic order can be suppressed by an in-plane magnetic field, and the field-induced spin disordered state is proposed to be Kitaev spin liquid (KSL) \cite{rucl3f1,rucl3f2,rucl3f3,rucl3f4,rucl3f5,rucl3f6}. 

Kitaev spin liquids could harbor unconventional excitations, most notably Majorana fermions \cite{kitaev}. The signature of the Majorana mode can be detected through scattering and transport experiments \cite{rclneutron,rucl31,rucl32,th1,th2}. Thermal transport experiment is uniquely suited for the study of these charge-neutral quasiparticles. There are two key findings from thermal transport studies on $\alpha$-RuCl$_3$. Firstly, the in-plane thermal Hall conductivity could reach half quantum thermal conductance ($\pi^2 k_B^2 / 3 h$) \cite{th1,th2,th3}. Secondly, the thermal conductivity shows oscillatory features as a function of in-plane magnetic field \cite{thos1,thos2}. These two key findings reflect the dual aspects of the low energy excitation gap in $\alpha$-RuCl$_3$. On one hand, the in-plane thermal Hall effect suggests that $\alpha$-RuCl$_3$ should host gapped bands with opposite topological indices \cite{th1}. On the other hand, the quantum oscillation interpretation \cite{thos1} implies the existence of a Fermi surface, and therefore gapless fermionic excitations. The physical origin of these observations is still under debate \cite{th1,th2,th3,th4,thos1,thos2,thos3}.


For the Kitaev model, the direction of the magnetic field can be consequential. For example, the Majorana fermion acquires a topological gap only for magnetic field perpendicular to the honeycomb bond \cite{kitaev}. And braiding is well-defined only for gapped, localizable excitations \cite{kitaev}. Recent field-angle dependent studies of $\alpha$-RuCl$_3$ using different techniques provide important insights on this candidate KSL material. For example, the angular dependent heat capacity measurement suggests the existence of a fermionic excitation, whose spectral gap evolves with magnetic field vector in the same way as predicted by the Kitaev model \cite{rucl3cp}. Furthermore, angular dependent thermal transport studies show that the in-plane thermal Hall effect has the sign-structure which follows prediction by the Kitaev interactions \cite{supp,ang3}. In this context (thermal Hall and oscillations), the natural following step is to investigate the angular dependence of the oscillatory features of thermal conductivity in $\alpha$-RuCl$_3$. This is the focus of our current work.

In this paper, we focus on the oscillatory features of thermal conductivity and investigate its anisotropic properties. With fixed field directions (B//a or B//b), we measured the thermal conductivity with thermal currents along a-axis (J//a) and b-axis (J//b). We found that the oscillatory features for orthogonal thermal currents (ka and kb) behave in-sync when the magnetic field is along the a-axis (B//a), and out-of-sync when the magnetic field is along the b-axis (B//b). We discuss the relevancy of existing interpretations (specifically, quantum oscillation \cite{thos1} and phase transition \cite{thos2}) to our data and propose our understanding on the matter. Furthermore, we investigate the oscillatory features' evolution as the magnetic field is switched between B//b to B//a. We found that the oscillatory features show distinct rotational symmetry before and after the emergence of the Kitaev spin liquid phase, showing higher order oscillations (doubling of angular frequency) as KSL phase emerges. Our work shows that in all respects, the oscillatory features of thermal conductivity in $\alpha$-RuCl$_3$ seem to closely tie to the Kitaev spin liquid phase, and the corresponding charge neutral fermion it holds.

\begin{figure} \centering \includegraphics [width = 0.4\textwidth] {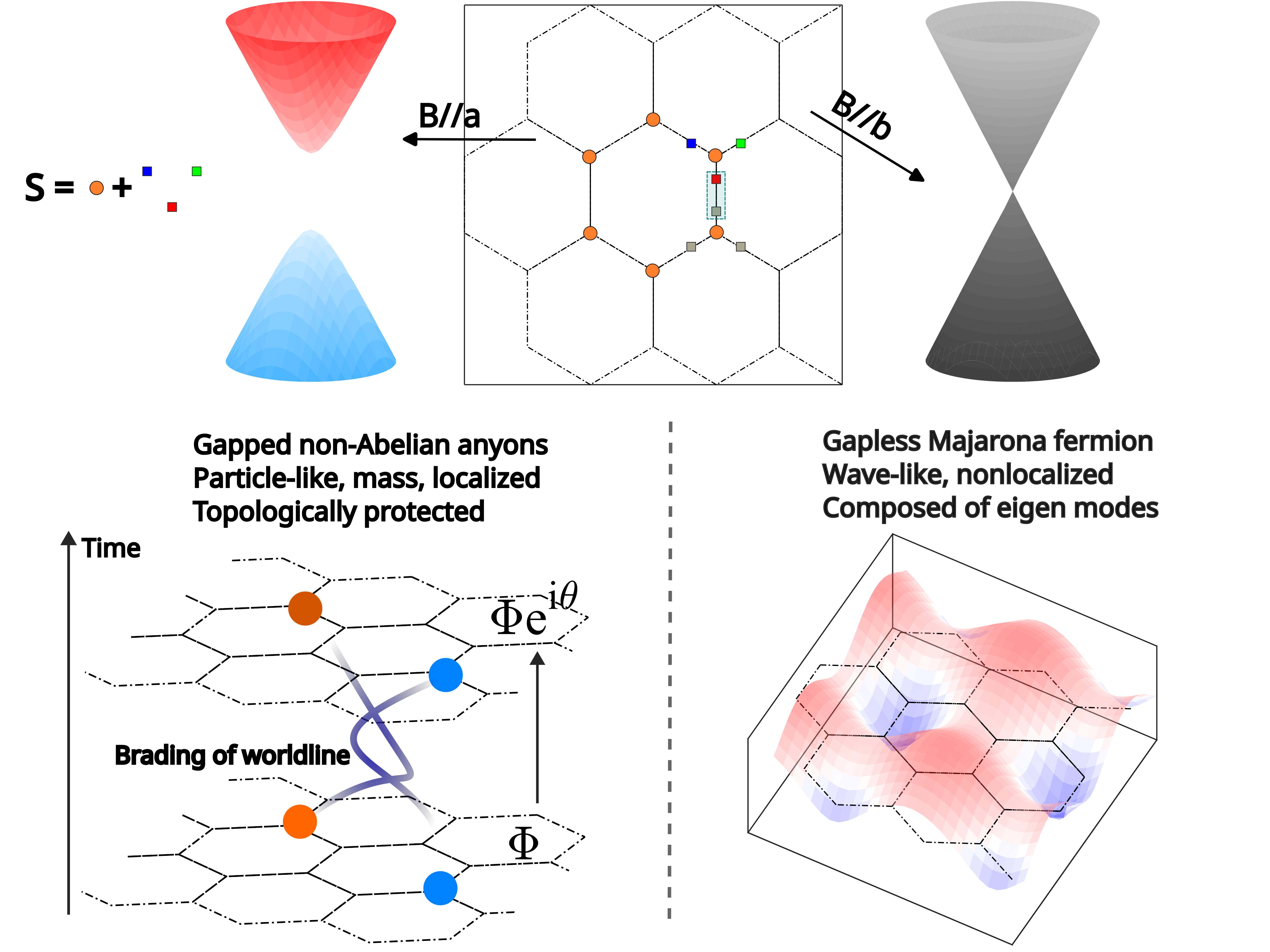}
\caption{Kitaev model and the spectrum of the free Majorana fermion. A spin operator fractionalized to one free Majorana fermion (orange) and three local gauge Majorana fermion (blue, green and red). The mutual statistics of the free Majorana fermion and the gauge Majorana fermions effectively presents a model of anyon. The braiding trajectory of the world lines of the free Majorana fermion encodes information as the phase change of wave functions, a mechanism proposed to be used in topological quantum computation. When the spectrum is gapless, the free Majorana fermion cannot be localized, and braiding is not well defined. In $\alpha$-RuCl$_3$, the two cases are simply thirty degrees away, as the magnetic field is switched between B//a (gapped) and B//b (gapless).}
\label{fig1}
\end{figure}


Single crystals with minimal secondary magnetic phases were used in this work \cite{supp}. The field dependent thermal conductivity and the oscillatory features measured at T = 2 K are shown in Figure 2. We investigated a total of four experimental configurations on two samples: $\Vec{J} \parallel \Vec{B} \parallel \Vec{a}$, $\Vec{J} \parallel \Vec{B}  \parallel \Vec{b}$, $\Vec{J} \perp \Vec{B} \parallel \Vec{a}$ and $\Vec{J} \perp \Vec{B} \parallel \Vec{b}$. Here, $\Vec{B}$, $\Vec{J}$ and $\Vec{a}$/$\Vec{b}$ represents the magnetic field vector, the thermal current vector, and the crystal axis perpendicular/parallel to Ru-Ru bond. The four curves shown in Figure 2(a, c) can generally be viewed as fine oscillatory features [Figure 2(b, d)] superimposed upon an oscillation-free background \cite{supp}. The background has a broad valley in the region where the Kitaev spin liquid phase dominate (light orange shaded). The interpretation of this valley is that phonon and other low energy spin excitations in the system coexist and scatter strongly in this region, suppressing the thermal conductivity \cite{th1,th2,th3,th4,thos1,thos2,thos3}. In contrast to the background, the amplitudes of the oscillatory features are strongly enhanced within this region, as shown in Figure 2(b, d). Focusing on this region, we take the normalized first derivative ($k^{-1}dk/dB$) and accentuate the oscillatory features, as shown in Figure 2(b, d). We find that the oscillatory features of k$_a$ and k$_b$ are mainly in pace with each other for B//a (gapped), whereas they are out-of-sync for B//b (gapless). 

\begin{figure*} \centering \includegraphics [width = 0.85\textwidth] {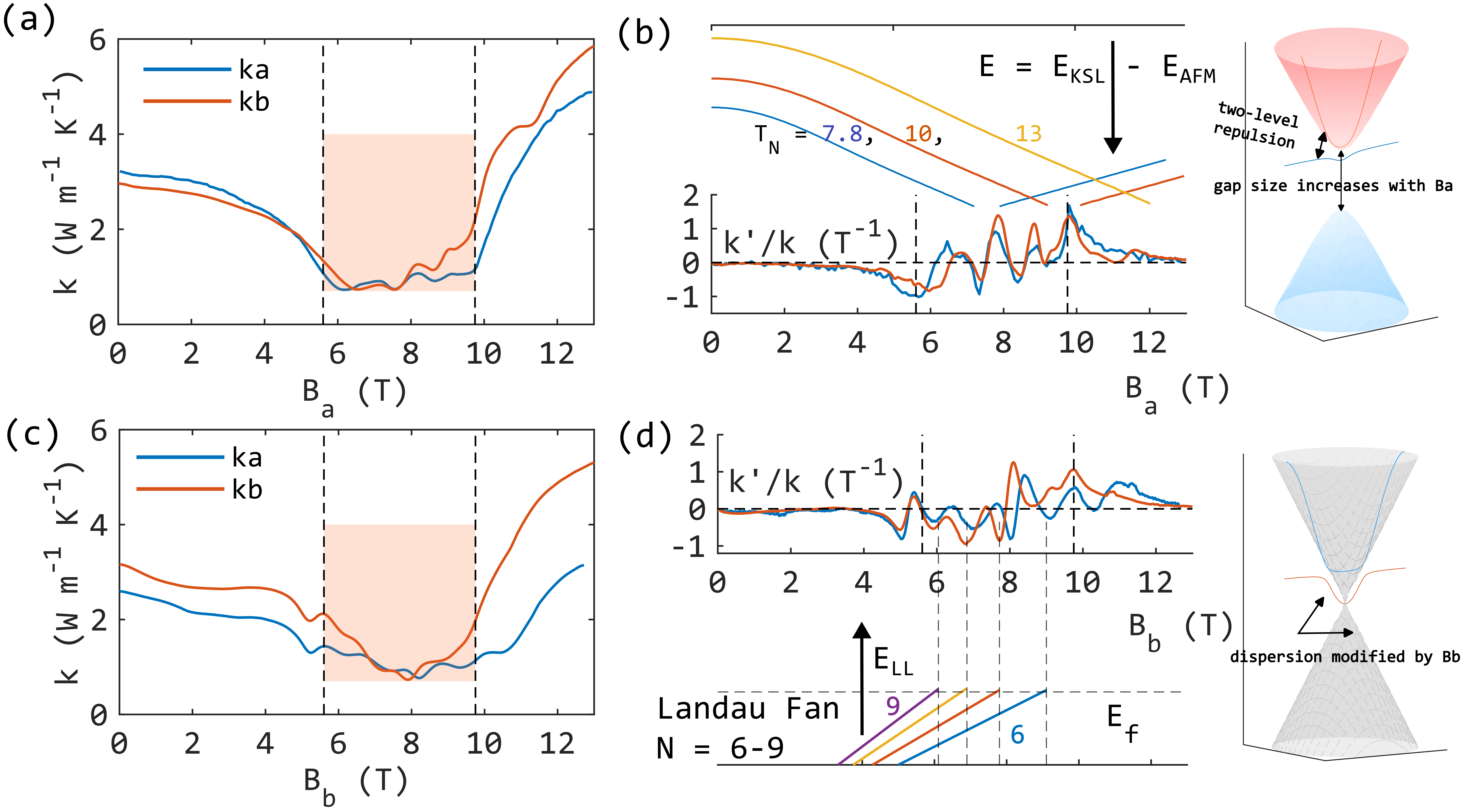}
\caption{Field dependent thermal conductivity (ka, kb) and its oscillatory features ($k^{'}/k$) at T = 2 K, measured with the magnetic field applied along the a-axis (a, b) and b-axis (c, d). Here, a/b refers to perpendicular/parallel to Ru-Ru bond, respectively. Illustrations of the two popular interpretations of the oscillatory feature are shown in (b, d), see main text. Neither interpretation provide a satisfactory explanation to experiment data. The proposed mechanism to explain the difference in oscillatory features for B//a and B//b, in terms of the charge neutral fermion and its interaction with other low energy excitation in $\alpha$-RuCl$_3$.}
\label{fig2}
\end{figure*}

The oscillatory features of thermal conductivity in $\alpha-$RuCl$_3$ is not an ordinary phenomenon. Currently, there are two main proposals for explaining this observation. The first is to draw an analogy to the Shubnikov–de Haas (SdH) effect \cite{thos1}. For the SdH effect, resistivity oscillates with magnetic field as Landau levels pass through the Fermi energy [Figure 2(d)]. The oscillation stops when the lowest Landau level has passed the Fermi energy (i.e., in the extreme quantum limit). The second explanation for the oscillatory features is that they are signatures of field-induced transitions due to secondary magnetic phases \cite{thos2}. In other words, secondary magnetic phases with different Neel temperatures are suppressed by different field strength, mimicking quantum oscillation behavior [Figure 2(b)]. 

Let us examine both hypothesis against the oscillatory feature data in Figure 2 (b, d). In the quantum oscillation picture, one should intuitively expect the oscillatory features to depend on both the field vector $\Vec{B}$ and the thermal current vector $\Vec{J}$. This is because the conductivity tensor will depend on both the density of states and the crystal momentum. If we naively apply the Landau quantization mechanism ($E_{LL} = (N+1/2)*\hbar qB/m$) and test it against data in Figure 2(d) (B//b, gapless), we could obtain reasonable prediction for the critical fields by setting appropriate Fermi energy \cite{supp}. It is worth noting that while the oscillation amplitude is sample-dependent, the critical fields for the oscillations agree reasonably well between results from different groups \cite{thos1,thos2,thos3}. However, for B//a, a gapped spectrum seemingly suggests that there should not be a Fermi surface, nor quantum oscillation. In the second scenario, the energy difference between the spin liquid state and the magnetic order state ($\Delta E = E_{KSL} - E_{AFM}$) decreases as the field strength increases. At each critical field where a long-range magnetic order is suppressed (determined by the Neel ordering temperature), there is a strong suppression of thermal conductivity due to this phase transition. For the second scenario to hold true, we expect the oscillatory features to depend only on the field vector $\Vec{B}$. This is because near phase transitions, strong fluctuations occur in every direction, thus k$_a$ and k$_b$ should oscillate in pace with each other. The observation of in-sync ka and kb oscillations for B//a is in line with this picture. However, the out-of-sync behavior of ka and kb oscillations for B//b seem to contradict this interpretation.

Incomplete agreement between the proposed physical pictures discussed above and the experiment data is an indication that a broader framework of thinking is necessary. As we draw an analogy from the SdH effect, we also unintentionally carry over a false assumption: the assumption that only the (spinon) Fermi surface matters for the oscillatory features of (thermal) conductivity. This is mostly true for the SdH effect. First, only charged particles (electrons) can contribute to electric resistivity of a system. Second, the scale of the fermi energy (of electrons) is much higher than any other excitations in the system (e.g., phonons, magnons, etc). The first condition ensures that only charge fermions need to be considered, and the second condition ensures that the charge fermions' interaction with other quasiparticles can be omitted, since they are buried deep in the fermi sea. Neither condition is satisfied here for $\alpha$-RuCl$_3$. First, recall that thermal conductivity is a measure of the system's ability to carry heat, and all quasiparticles which can carry thermal energy in principle contribute to this physical measurable. Second, the energy scale of the spinon (meV) is on a comparable level to phonon and magnon, etc. The interactions between spinons and other quasiparticles in the system can greatly modify the spinon spectrum. Therefore, as we consider the oscillatory features of thermal conductivity in $\alpha$-RuCl$_3$, the spinon Fermi surface does not stand in isolation. Instead, its own geometry, as well as its interactions with other low energy excitations should be considered.


In this context, we can re-examine the data in Figure 2 with a broader perspective. First point of consideration is why does thermal conductivity oscillate? Both mechanisms discussed previously describe a process where a series of levels ($E_{LL}$ or $\Delta E$) pass through a thresh-hold value ($E_{f}$ or 0) as magnetic field changes, causing the thermal conductivity to oscillate. In principle, the threshold values could also be the energy where the spinon band and other lower energy excitation branch meet. We can consider a two-level Hamiltonian near this band crossing. Adding an interaction term cause a repulsion between the level, modifying both bands. This is a simple yet general model widely applicable to many magnetic systems. A second point of considerations is why do the oscillatory features for ka/kb behave differently in the presences of Ba/Bb? The main difference between Ba/Bb is the presence/absence of a spectral gap for the charge neutral free fermion \cite{rucl3cp}. For B//a, field strength acts to increase the gap size \cite{rucl3cp}. Peaks/valleys of thermal conductivity comes about as the charge neutral fermion and other lower energy excitations come in and out of contact in reciprocal space. This process is isotropic with respect to the crystal momentum, thus isotropic response for $k_a$ and $k_b$ is anticipated. The exact physical picture is complicated: various spectroscopic methods have identified many different low energy excitations in $\alpha$-RuCl$_3$. The low energy excitation could be phonon \cite{phone}, magnon and their bound states \cite{meg1,meg2}, or even continuum spin excitations \cite{conti1,conti2}. However, in terms of thermal transport properties, the fermionic excitation is the primary consideration: whether it is the in-plane thermal Hall effect (Majorana fermion) \cite{th1}, oscillatory feature of thermal conductivity (spinon) \cite{thos1} and its anisotropic behavior studied here. 


\begin{figure*} \centering \includegraphics [width = 0.85\textwidth] {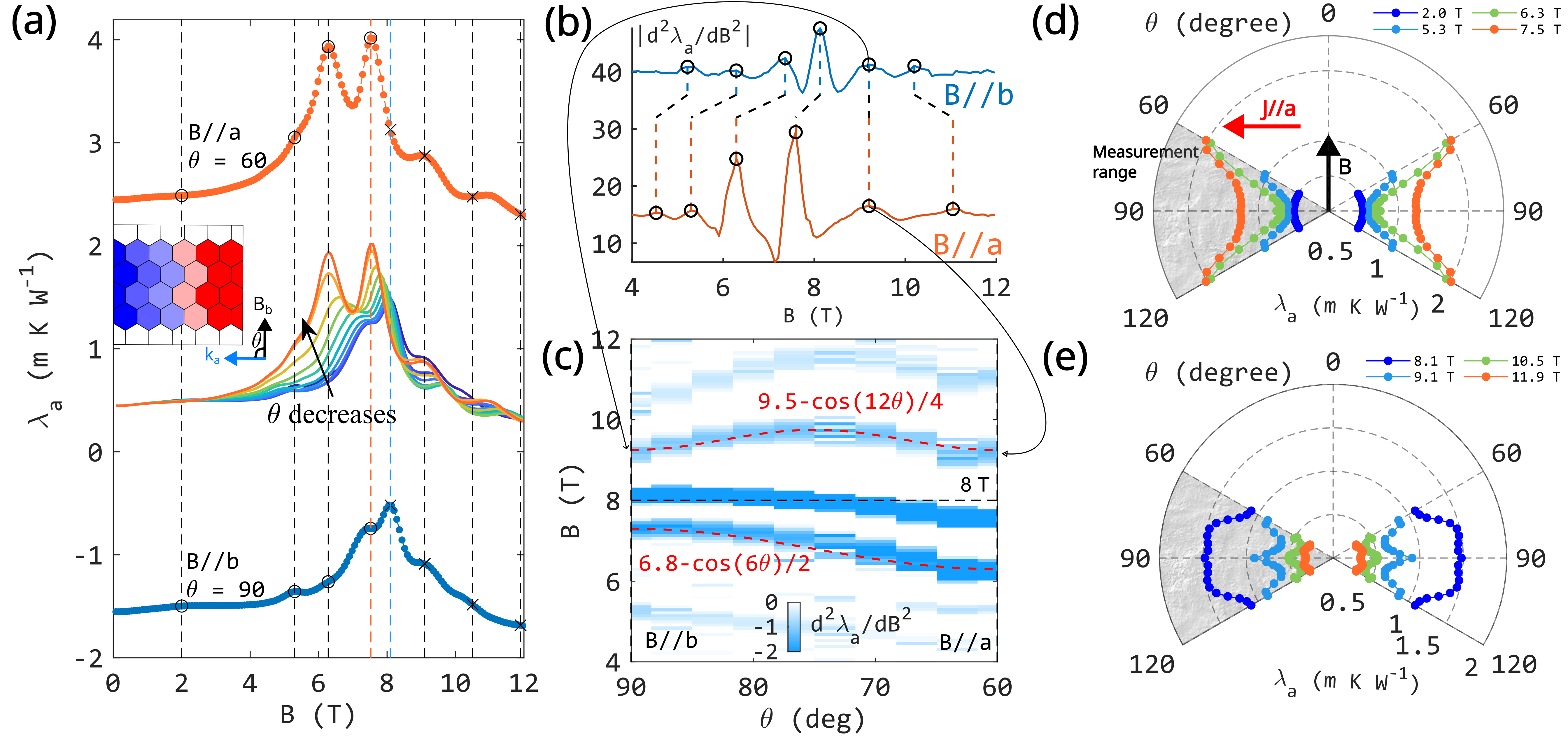}
\caption{(a) Thermal resistivity ($\lambda_a$) measured with $\theta$ from 90$\degree$ to 60$\degree$ at $d\theta \approx$ 3.33 $\degree$ per step. The curves for B//a (orange) and B//b (blue) have been shifted vertical by $2 \ W m^{-1} K^{-1}$ for clarity. (Inset) Illustration of the experiment setup at $\theta$ = 90$\degree$ (B//b). (b) Second field derivative of $\lambda_a$. Black circle markers indicate the peaks, and black dash lines indicates the one-to-one correspondence as field switch from B//a to B//b. (c) Pseudo-color map of the second field derivative of $\lambda_a$. The two cosine functions (red dash) are intended for guide-to-eye only. (d, e) Polar plot of $\lambda_a$ at selected fields showing different rotational symmetry (d) before and (e) after the emergence of the Kitaev spin liquid phase. $\theta$ denotes angle between $\Vec{B}$ to $\Vec{J}$. Gray area indicates measurement range. Additional range is generated by assuming the symmetry constraint that $\lambda_a (\Vec{J}) = \lambda_a (-\Vec{J})$ \cite{supp}.}
\label{fig3}
\end{figure*}

The magnetic field as a vector, governs the underlying physics of $\alpha$-RuCl$_3$. From a theoretical viewpoint, different field directions determine whether we are studying a non-Abelian anyon or a gapless Majorana fermion. From an experimental viewpoint, both the in-plane thermal Hall effect and oscillatory features of thermal conductivity behave distinctly different for B//a and B//b. A natural question is how are the gapped (B//a) and gapless (B//b) cases connected as the magnetic field rotates? How does the thermal transport properties evolve in response to changes of the magnetic field direction?

Figure 3 (a) shows the thermal resistivity measured with J//a ($\lambda_a$), and the magnetic field applied at $\theta$ between $90 \degree$ (B//b) and $60 \degree$ (B//a) at about $3.33 \degree$ per step. Here, $\theta$ refers to the relative angle between $\Vec{B}$ and $\Vec{J}$. Inset shows an illustration of the experiment setup for $\theta = 90 \degree$ (B//b). In the field range B = 4 T to B = 8 T, $\lambda_a$ decreases monotonically as $\theta$ increases. The oscillatory features at B//b and B//a are composed of the same set of peaks as the field rotates. In Figure 3(b), the oscillatory feature ($|d^{2} \lambda_a/dB^{2}|$) at the two extremes (B//a and B//b) are shown, where we could identify a one-to-one correspondence for peaks indicated by the black circle marks \cite{supp}. In Figure 3(c), we track the evolution of this correspondence by constructing the pseudo-color map of the oscillatory features as functions of field strength and angle ($\theta$). Peaks of oscillatory features are the blue regions, weaker oscillatory features are hard to discern due to insufficient contrast. At approximately 8 T, the field is sufficiently strong to induce the phase transition from conventional order to Kitaev spin liquid state, irrespective of the field direction. The angular-dependence of the oscillatory features are different for field strengths above and below 8 T. The red dash lines in Figure 3(c) are cosine functions intended to roughly trace out the peak-$\theta$ dependence. For oscillatory features below 8 T, the peak-$\theta$ dependence is monotonic. The $\theta$-frequency for oscillatory features above 8 T is nearly doubled, compared to data below 8 T. For example, the peak near 9.25 T (red dash) are nearly at the same critical field for B//a and B//b, while it reaches its maximum peak location (9.75 T) near $\theta = 75 \degree$.

Several characteristic fields (where $\lambda_{a}$ peaks) were chosen to produce the angular dependent polar plots shown in Figure 3(d, e). The vertical dash lines in Figure 3(a) indicates field values which were chosen. For field strength below 8 T (d), we observe that $\lambda_{a}$ gradually evolves from being principally isotropic (2T, circular) to principally six-fold anisotropic (7.5 T, \cite{supp}). For field strength above 8 T (e), there is a doubling of angular-frequency of thermal resistivity. At high fields (above 12 T) where the system nearly enters the polarized state, the anisotropic character of $\lambda_{a}$ is gradually suppressed. The oscillatory features of $\lambda_{a}$ shows distinct $\theta$-dependence before and after the emergence of Kitaev spin liquid phase. Except for the very low ($<$ 2 T) and high field regions ($>$ 12 T), $\lambda_{a}$ is highly anisotropic. This indicates that while phonons contribute to the thermal conductivity, it alone cannot dictate the rotational symmetry of $\lambda_{a}$ and its oscillatory features. For example, at B = 2 T where phonon contribution dominates, $\lambda_{a}$ is nearly isotropic with respect to $\theta$. At intermediate field strength, $\lambda_{a}$ shows six-fold symmetry \cite{supp}, similar to the six-fold symmetry observed in the thermal dynamic study \cite{rucl3cp}. This is unsurprising since the Kitaev model is an anisotropic spin model on a honeycomb lattice (six-fold). The doubling of angular-frequency for $\lambda_{a}$  above 8 T is unexpected. 

The symmetry of transport coefficients reflects the symmetry of the physical mechanism responsible for them. Vice versa, identifying new symmetries of the transport coefficients could lead to the discovery of new states/physical mechanisms. For example, six-fold in-plane rotational symmetry of the second critical field ($H_{c2}$) of an Ising superconductor signifies finite-momentum Copper pairing in 2H-HbSe$_2$ (Fulde-Ferrell-Larkin-Ovchinnikov state) \cite{fflo}. The planar anisotropic magnetoresistivity of a topological kagome metal KV$_3$Sb$_5$ shows higher-order oscillation (doubling of angular-frequency) above 10 T \cite{kvs}. Both instances emphasize on the rotational symmetry change of the transport coefficient as evidence for either the emergence of new states or the presence of strong correlations. 

The generic in-plane magneto-resistance effect depends only on the angle between $\Vec{B}$ and $\Vec{J}$ (i.e., $\Delta \rho \propto cos^{2}(\theta)$ \cite{mr}), and deviation from this two-fold symmetry has only been observed in films or nano-flakes \cite{Bifilm,laosto,kvs}. In these cases, the magnetic field either directly modifies the Fermi surface (Bi thin film, \cite{Bifilm}), or indirectly affects $\rho$ through magnetic scattering (LAO/STO interface, \cite{laosto}). The doubling of angular frequency for $\lambda_{a}$ and its oscillatory feature may have similar implications here. In $\alpha$-RuCl$_3$, there is also a Fermi surface involved, except now it is that of a charge neutral fermion. While neither spinons nor Majorana fermions carry charges, they both carry spin (S=1/2), and therefore the magnetic field vector could directly modify the corresponding Fermi surface. Furthermore, the doubling of angular frequency for $\lambda_{a}$ only occurs as the Kitaev spin liquid phase emerges. One finds a similar situation in KV$_3$Sb$_5$: where the higher order oscillation of $\rho$ occurs simultaneously as the charge density wave (CDW) order establishes \cite{kvs}. The synchronous emergence of higher order oscillation and a new phase, either classical such as a CDW, or topological such as the Kitaev spin liquid phase, suggests that there should be a close correlation between the two. The exact physical mechanism is a subject warranted for future studies.

In conclusion, we investigate the oscillatory features of thermal conductivity in $\alpha$-RuCl$_3$, as functions of the thermal current direction and magnetic field vector. Depending on the filed direction, the thermal conductivity for orthogonal thermal current directions (ka, kb) either oscillates in-sync (B//a, gapped) or out-of-sync (B//b, gapless). We discuss the relevance of the charge-neutral fermion, its spectral gap and its interactions with other quasiparticles to the aforementioned observations. We also track the evolution of oscillatory features as functions of $\theta$, the angle between thermal current vector and magnetic field vector. The thermal resistivity shows a higher order oscillation (with respect to $\theta$) as the Kitaev spin liquid phase emerges. We point out the close resemblance of higher order oscillation of transport coefficients observed in $\alpha$-RuCl$_3$ and other two-dimensional electron systems: both tied to the magnetic field modification of a Fermi surface. 

\section{Acknowledgment}
HZ, SN, MM, and JY were supported by the U.S. Department of Energy, Office of Science, National Quantum Information Science Research Centers, Quantum Science Center. HM, DG and TW were supported by the US Department of Energy, Office of Science, Basic Energy Sciences, Materials Sciences and Engineering Division. 

This manuscript has been authored by UT-Battelle, LLC, under Contract No. DE-AC0500OR22725 with the U.S. Department of Energy. The United States Government retains and the publisher, by accepting the article for publication, acknowledges that the United States Government retains a non-exclusive, paid-up, irrevocable, world-wide license to publish or reproduce the published form of this manuscript, or allow others to do so, for the United States Government purposes. The Department of Energy will provide public access to these results of federally sponsored research in accordance with the DOE Public Access Plan (http://energy.gov/downloads/doe-public-access-plan).

\bibliographystyle{apsrev4-1} 
\bibliography{ref}

\begin{thebibliography}{34}%
\makeatletter
\providecommand \@ifxundefined [1]{%
 \@ifx{#1\undefined}
}%
\providecommand \@ifnum [1]{%
 \ifnum #1\expandafter \@firstoftwo
 \else \expandafter \@secondoftwo
 \fi
}%
\providecommand \@ifx [1]{%
 \ifx #1\expandafter \@firstoftwo
 \else \expandafter \@secondoftwo
 \fi
}%
\providecommand \natexlab [1]{#1}%
\providecommand \enquote  [1]{``#1''}%
\providecommand \bibnamefont  [1]{#1}%
\providecommand \bibfnamefont [1]{#1}%
\providecommand \citenamefont [1]{#1}%
\providecommand \href@noop [0]{\@secondoftwo}%
\providecommand \href [0]{\begingroup \@sanitize@url \@href}%
\providecommand \@href[1]{\@@startlink{#1}\@@href}%
\providecommand \@@href[1]{\endgroup#1\@@endlink}%
\providecommand \@sanitize@url [0]{\catcode `\\12\catcode `\$12\catcode
  `\&12\catcode `\#12\catcode `\^12\catcode `\_12\catcode `\%12\relax}%
\providecommand \@@startlink[1]{}%
\providecommand \@@endlink[0]{}%
\providecommand \url  [0]{\begingroup\@sanitize@url \@url }%
\providecommand \@url [1]{\endgroup\@href {#1}{\urlprefix }}%
\providecommand \urlprefix  [0]{URL }%
\providecommand \Eprint [0]{\href }%
\providecommand \doibase [0]{http://dx.doi.org/}%
\providecommand \selectlanguage [0]{\@gobble}%
\providecommand \bibinfo  [0]{\@secondoftwo}%
\providecommand \bibfield  [0]{\@secondoftwo}%
\providecommand \translation [1]{[#1]}%
\providecommand \BibitemOpen [0]{}%
\providecommand \bibitemStop [0]{}%
\providecommand \bibitemNoStop [0]{.\EOS\space}%
\providecommand \EOS [0]{\spacefactor3000\relax}%
\providecommand \BibitemShut  [1]{\csname bibitem#1\endcsname}%
\let\auto@bib@innerbib\@empty
\bibitem [{\citenamefont {Kitaev}(2006)}]{kitaev}%
  \BibitemOpen
  \bibfield  {author} {\bibinfo {author} {\bibfnamefont {A.}~\bibnamefont
  {Kitaev}},\ }\href {\doibase https://doi.org/10.1016/j.aop.2005.10.005}
  {\bibfield  {journal} {\bibinfo  {journal} {Annals of Physics}\ }\textbf
  {\bibinfo {volume} {321}},\ \bibinfo {pages} {2} (\bibinfo {year} {2006})},\
  \bibinfo {note} {january Special Issue}\BibitemShut {NoStop}%
\bibitem [{\citenamefont {Kitaev}(2003)}]{kitaevqc}%
  \BibitemOpen
  \bibfield  {author} {\bibinfo {author} {\bibfnamefont {A.}~\bibnamefont
  {Kitaev}},\ }\href {\doibase https://doi.org/10.1016/S0003-4916(02)00018-0}
  {\bibfield  {journal} {\bibinfo  {journal} {Annals of Physics}\ }\textbf
  {\bibinfo {volume} {303}},\ \bibinfo {pages} {2} (\bibinfo {year}
  {2003})}\BibitemShut {NoStop}%
\bibitem [{\citenamefont {Plumb}\ \emph {et~al.}(2014)\citenamefont {Plumb},
  \citenamefont {Clancy}, \citenamefont {Sandilands}, \citenamefont {Shankar},
  \citenamefont {Hu}, \citenamefont {Burch}, \citenamefont {Kee},\ and\
  \citenamefont {Kim}}]{plum}%
  \BibitemOpen
  \bibfield  {author} {\bibinfo {author} {\bibfnamefont {K.}~\bibnamefont
  {Plumb}}, \bibinfo {author} {\bibfnamefont {J.}~\bibnamefont {Clancy}},
  \bibinfo {author} {\bibfnamefont {L.}~\bibnamefont {Sandilands}}, \bibinfo
  {author} {\bibfnamefont {V.~V.}\ \bibnamefont {Shankar}}, \bibinfo {author}
  {\bibfnamefont {Y.}~\bibnamefont {Hu}}, \bibinfo {author} {\bibfnamefont
  {K.}~\bibnamefont {Burch}}, \bibinfo {author} {\bibfnamefont {H.-Y.}\
  \bibnamefont {Kee}}, \ and\ \bibinfo {author} {\bibfnamefont {Y.-J.}\
  \bibnamefont {Kim}},\ }\href@noop {} {\bibfield  {journal} {\bibinfo
  {journal} {Phys. Rev. B}\ }\textbf {\bibinfo {volume} {90}},\ \bibinfo
  {pages} {041112} (\bibinfo {year} {2014})}\BibitemShut {NoStop}%
\bibitem [{\citenamefont {Banerjee}\ \emph {et~al.}(2016)\citenamefont
  {Banerjee}, \citenamefont {Bridges}, \citenamefont {Yan}, \citenamefont
  {Aczel}, \citenamefont {Li}, \citenamefont {Stone}, \citenamefont {Granroth},
  \citenamefont {Lumsden}, \citenamefont {Yiu}, \citenamefont {Knolle} \emph
  {et~al.}}]{banj}%
  \BibitemOpen
  \bibfield  {author} {\bibinfo {author} {\bibfnamefont {A.}~\bibnamefont
  {Banerjee}}, \bibinfo {author} {\bibfnamefont {C.}~\bibnamefont {Bridges}},
  \bibinfo {author} {\bibfnamefont {J.-Q.}\ \bibnamefont {Yan}}, \bibinfo
  {author} {\bibfnamefont {A.}~\bibnamefont {Aczel}}, \bibinfo {author}
  {\bibfnamefont {L.}~\bibnamefont {Li}}, \bibinfo {author} {\bibfnamefont
  {M.}~\bibnamefont {Stone}}, \bibinfo {author} {\bibfnamefont
  {G.}~\bibnamefont {Granroth}}, \bibinfo {author} {\bibfnamefont
  {M.}~\bibnamefont {Lumsden}}, \bibinfo {author} {\bibfnamefont
  {Y.}~\bibnamefont {Yiu}}, \bibinfo {author} {\bibfnamefont {J.}~\bibnamefont
  {Knolle}},  \emph {et~al.},\ }\href@noop {} {\bibfield  {journal} {\bibinfo
  {journal} {Nat. Mater.}\ }\textbf {\bibinfo {volume} {15}},\ \bibinfo {pages}
  {733} (\bibinfo {year} {2016})}\BibitemShut {NoStop}%
\bibitem [{\citenamefont {Ran}\ \emph {et~al.}(2017)\citenamefont {Ran},
  \citenamefont {Wang}, \citenamefont {Wang}, \citenamefont {Dong},
  \citenamefont {Ren}, \citenamefont {Bao}, \citenamefont {Li}, \citenamefont
  {Ma}, \citenamefont {Gan}, \citenamefont {Zhang}, \citenamefont {Park},
  \citenamefont {Deng}, \citenamefont {Danilkin}, \citenamefont {Yu},
  \citenamefont {Li},\ and\ \citenamefont {Wen}}]{rucl31}%
  \BibitemOpen
  \bibfield  {author} {\bibinfo {author} {\bibfnamefont {K.}~\bibnamefont
  {Ran}}, \bibinfo {author} {\bibfnamefont {J.}~\bibnamefont {Wang}}, \bibinfo
  {author} {\bibfnamefont {W.}~\bibnamefont {Wang}}, \bibinfo {author}
  {\bibfnamefont {Z.-Y.}\ \bibnamefont {Dong}}, \bibinfo {author}
  {\bibfnamefont {X.}~\bibnamefont {Ren}}, \bibinfo {author} {\bibfnamefont
  {S.}~\bibnamefont {Bao}}, \bibinfo {author} {\bibfnamefont {S.}~\bibnamefont
  {Li}}, \bibinfo {author} {\bibfnamefont {Z.}~\bibnamefont {Ma}}, \bibinfo
  {author} {\bibfnamefont {Y.}~\bibnamefont {Gan}}, \bibinfo {author}
  {\bibfnamefont {Y.}~\bibnamefont {Zhang}}, \bibinfo {author} {\bibfnamefont
  {J.~T.}\ \bibnamefont {Park}}, \bibinfo {author} {\bibfnamefont
  {G.}~\bibnamefont {Deng}}, \bibinfo {author} {\bibfnamefont {S.}~\bibnamefont
  {Danilkin}}, \bibinfo {author} {\bibfnamefont {S.-L.}\ \bibnamefont {Yu}},
  \bibinfo {author} {\bibfnamefont {J.-X.}\ \bibnamefont {Li}}, \ and\ \bibinfo
  {author} {\bibfnamefont {J.}~\bibnamefont {Wen}},\ }\href {\doibase
  10.1103/PhysRevLett.118.107203} {\bibfield  {journal} {\bibinfo  {journal}
  {Phys. Rev. Lett.}\ }\textbf {\bibinfo {volume} {118}},\ \bibinfo {pages}
  {107203} (\bibinfo {year} {2017})}\BibitemShut {NoStop}%
\bibitem [{\citenamefont {Do}\ \emph {et~al.}(2017)\citenamefont {Do},
  \citenamefont {Park}, \citenamefont {Yoshitake}, \citenamefont {Nasu},
  \citenamefont {Motome}, \citenamefont {Kwon}, \citenamefont {Adroja},
  \citenamefont {Voneshen}, \citenamefont {Kim}, \citenamefont {Jang},
  \citenamefont {Park}, \citenamefont {Choi},\ and\ \citenamefont
  {Ji}}]{rucl32}%
  \BibitemOpen
  \bibfield  {author} {\bibinfo {author} {\bibfnamefont {S.-H.}\ \bibnamefont
  {Do}}, \bibinfo {author} {\bibfnamefont {S.-Y.}\ \bibnamefont {Park}},
  \bibinfo {author} {\bibfnamefont {J.}~\bibnamefont {Yoshitake}}, \bibinfo
  {author} {\bibfnamefont {J.}~\bibnamefont {Nasu}}, \bibinfo {author}
  {\bibfnamefont {Y.}~\bibnamefont {Motome}}, \bibinfo {author} {\bibfnamefont
  {Y.}~\bibnamefont {Kwon}}, \bibinfo {author} {\bibfnamefont {D.~T.}\
  \bibnamefont {Adroja}}, \bibinfo {author} {\bibfnamefont {D.~J.}\
  \bibnamefont {Voneshen}}, \bibinfo {author} {\bibfnamefont {K.}~\bibnamefont
  {Kim}}, \bibinfo {author} {\bibfnamefont {T.-H.}\ \bibnamefont {Jang}},
  \bibinfo {author} {\bibfnamefont {J.-H.}\ \bibnamefont {Park}}, \bibinfo
  {author} {\bibfnamefont {K.-Y.}\ \bibnamefont {Choi}}, \ and\ \bibinfo
  {author} {\bibfnamefont {S.}~\bibnamefont {Ji}},\ }\href {\doibase
  10.1038/nphys4264} {\bibfield  {journal} {\bibinfo  {journal} {Nature
  Physics}\ }\textbf {\bibinfo {volume} {13}},\ \bibinfo {pages} {1079}
  (\bibinfo {year} {2017})}\BibitemShut {NoStop}%
\bibitem [{\citenamefont {Cao}\ \emph {et~al.}(2016)\citenamefont {Cao},
  \citenamefont {Banerjee}, \citenamefont {Yan}, \citenamefont {Bridges},
  \citenamefont {Lumsden}, \citenamefont {Mandrus}, \citenamefont {Tennant},
  \citenamefont {Chakoumakos},\ and\ \citenamefont {Nagler}}]{rucl33}%
  \BibitemOpen
  \bibfield  {author} {\bibinfo {author} {\bibfnamefont {H.~B.}\ \bibnamefont
  {Cao}}, \bibinfo {author} {\bibfnamefont {A.}~\bibnamefont {Banerjee}},
  \bibinfo {author} {\bibfnamefont {J.-Q.}\ \bibnamefont {Yan}}, \bibinfo
  {author} {\bibfnamefont {C.~A.}\ \bibnamefont {Bridges}}, \bibinfo {author}
  {\bibfnamefont {M.~D.}\ \bibnamefont {Lumsden}}, \bibinfo {author}
  {\bibfnamefont {D.~G.}\ \bibnamefont {Mandrus}}, \bibinfo {author}
  {\bibfnamefont {D.~A.}\ \bibnamefont {Tennant}}, \bibinfo {author}
  {\bibfnamefont {B.~C.}\ \bibnamefont {Chakoumakos}}, \ and\ \bibinfo {author}
  {\bibfnamefont {S.~E.}\ \bibnamefont {Nagler}},\ }\href {\doibase
  10.1103/PhysRevB.93.134423} {\bibfield  {journal} {\bibinfo  {journal} {Phys.
  Rev. B}\ }\textbf {\bibinfo {volume} {93}},\ \bibinfo {pages} {134423}
  (\bibinfo {year} {2016})}\BibitemShut {NoStop}%
\bibitem [{\citenamefont {Sears}\ \emph {et~al.}(2017)\citenamefont {Sears},
  \citenamefont {Zhao}, \citenamefont {Xu}, \citenamefont {Lynn},\ and\
  \citenamefont {Kim}}]{rucl3f1}%
  \BibitemOpen
  \bibfield  {author} {\bibinfo {author} {\bibfnamefont {J.~A.}\ \bibnamefont
  {Sears}}, \bibinfo {author} {\bibfnamefont {Y.}~\bibnamefont {Zhao}},
  \bibinfo {author} {\bibfnamefont {Z.}~\bibnamefont {Xu}}, \bibinfo {author}
  {\bibfnamefont {J.~W.}\ \bibnamefont {Lynn}}, \ and\ \bibinfo {author}
  {\bibfnamefont {Y.-J.}\ \bibnamefont {Kim}},\ }\href {\doibase
  10.1103/PhysRevB.95.180411} {\bibfield  {journal} {\bibinfo  {journal} {Phys.
  Rev. B}\ }\textbf {\bibinfo {volume} {95}},\ \bibinfo {pages} {180411}
  (\bibinfo {year} {2017})}\BibitemShut {NoStop}%
\bibitem [{\citenamefont {Zheng}\ \emph {et~al.}(2017)\citenamefont {Zheng},
  \citenamefont {Ran}, \citenamefont {Li}, \citenamefont {Wang}, \citenamefont
  {Wang}, \citenamefont {Liu}, \citenamefont {Liu}, \citenamefont {Normand},
  \citenamefont {Wen},\ and\ \citenamefont {Yu}}]{rucl3f2}%
  \BibitemOpen
  \bibfield  {author} {\bibinfo {author} {\bibfnamefont {J.}~\bibnamefont
  {Zheng}}, \bibinfo {author} {\bibfnamefont {K.}~\bibnamefont {Ran}}, \bibinfo
  {author} {\bibfnamefont {T.}~\bibnamefont {Li}}, \bibinfo {author}
  {\bibfnamefont {J.}~\bibnamefont {Wang}}, \bibinfo {author} {\bibfnamefont
  {P.}~\bibnamefont {Wang}}, \bibinfo {author} {\bibfnamefont {B.}~\bibnamefont
  {Liu}}, \bibinfo {author} {\bibfnamefont {Z.-X.}\ \bibnamefont {Liu}},
  \bibinfo {author} {\bibfnamefont {B.}~\bibnamefont {Normand}}, \bibinfo
  {author} {\bibfnamefont {J.}~\bibnamefont {Wen}}, \ and\ \bibinfo {author}
  {\bibfnamefont {W.}~\bibnamefont {Yu}},\ }\href {\doibase
  10.1103/PhysRevLett.119.227208} {\bibfield  {journal} {\bibinfo  {journal}
  {Phys. Rev. Lett.}\ }\textbf {\bibinfo {volume} {119}},\ \bibinfo {pages}
  {227208} (\bibinfo {year} {2017})}\BibitemShut {NoStop}%
\bibitem [{\citenamefont {Hentrich}\ \emph {et~al.}(2018)\citenamefont
  {Hentrich}, \citenamefont {Wolter}, \citenamefont {Zotos}, \citenamefont
  {Brenig}, \citenamefont {Nowak}, \citenamefont {Isaeva}, \citenamefont
  {Doert}, \citenamefont {Banerjee}, \citenamefont {Lampen-Kelley},
  \citenamefont {Mandrus}, \citenamefont {Nagler}, \citenamefont {Sears},
  \citenamefont {Kim}, \citenamefont {B\"uchner},\ and\ \citenamefont
  {Hess}}]{rucl3f3}%
  \BibitemOpen
  \bibfield  {author} {\bibinfo {author} {\bibfnamefont {R.}~\bibnamefont
  {Hentrich}}, \bibinfo {author} {\bibfnamefont {A.~U.~B.}\ \bibnamefont
  {Wolter}}, \bibinfo {author} {\bibfnamefont {X.}~\bibnamefont {Zotos}},
  \bibinfo {author} {\bibfnamefont {W.}~\bibnamefont {Brenig}}, \bibinfo
  {author} {\bibfnamefont {D.}~\bibnamefont {Nowak}}, \bibinfo {author}
  {\bibfnamefont {A.}~\bibnamefont {Isaeva}}, \bibinfo {author} {\bibfnamefont
  {T.}~\bibnamefont {Doert}}, \bibinfo {author} {\bibfnamefont
  {A.}~\bibnamefont {Banerjee}}, \bibinfo {author} {\bibfnamefont
  {P.}~\bibnamefont {Lampen-Kelley}}, \bibinfo {author} {\bibfnamefont {D.~G.}\
  \bibnamefont {Mandrus}}, \bibinfo {author} {\bibfnamefont {S.~E.}\
  \bibnamefont {Nagler}}, \bibinfo {author} {\bibfnamefont {J.}~\bibnamefont
  {Sears}}, \bibinfo {author} {\bibfnamefont {Y.-J.}\ \bibnamefont {Kim}},
  \bibinfo {author} {\bibfnamefont {B.}~\bibnamefont {B\"uchner}}, \ and\
  \bibinfo {author} {\bibfnamefont {C.}~\bibnamefont {Hess}},\ }\href {\doibase
  10.1103/PhysRevLett.120.117204} {\bibfield  {journal} {\bibinfo  {journal}
  {Phys. Rev. Lett.}\ }\textbf {\bibinfo {volume} {120}},\ \bibinfo {pages}
  {117204} (\bibinfo {year} {2018})}\BibitemShut {NoStop}%
\bibitem [{\citenamefont {Wolter}\ \emph {et~al.}(2017)\citenamefont {Wolter},
  \citenamefont {Corredor}, \citenamefont {Janssen}, \citenamefont {Nenkov},
  \citenamefont {Sch\"onecker}, \citenamefont {Do}, \citenamefont {Choi},
  \citenamefont {Albrecht}, \citenamefont {Hunger}, \citenamefont {Doert},
  \citenamefont {Vojta},\ and\ \citenamefont {B\"uchner}}]{rucl3f4}%
  \BibitemOpen
  \bibfield  {author} {\bibinfo {author} {\bibfnamefont {A.~U.~B.}\
  \bibnamefont {Wolter}}, \bibinfo {author} {\bibfnamefont {L.~T.}\
  \bibnamefont {Corredor}}, \bibinfo {author} {\bibfnamefont {L.}~\bibnamefont
  {Janssen}}, \bibinfo {author} {\bibfnamefont {K.}~\bibnamefont {Nenkov}},
  \bibinfo {author} {\bibfnamefont {S.}~\bibnamefont {Sch\"onecker}}, \bibinfo
  {author} {\bibfnamefont {S.-H.}\ \bibnamefont {Do}}, \bibinfo {author}
  {\bibfnamefont {K.-Y.}\ \bibnamefont {Choi}}, \bibinfo {author}
  {\bibfnamefont {R.}~\bibnamefont {Albrecht}}, \bibinfo {author}
  {\bibfnamefont {J.}~\bibnamefont {Hunger}}, \bibinfo {author} {\bibfnamefont
  {T.}~\bibnamefont {Doert}}, \bibinfo {author} {\bibfnamefont
  {M.}~\bibnamefont {Vojta}}, \ and\ \bibinfo {author} {\bibfnamefont
  {B.}~\bibnamefont {B\"uchner}},\ }\href {\doibase 10.1103/PhysRevB.96.041405}
  {\bibfield  {journal} {\bibinfo  {journal} {Phys. Rev. B}\ }\textbf {\bibinfo
  {volume} {96}},\ \bibinfo {pages} {041405} (\bibinfo {year}
  {2017})}\BibitemShut {NoStop}%
\bibitem [{\citenamefont {Balz}\ \emph {et~al.}(2021)\citenamefont {Balz},
  \citenamefont {Janssen}, \citenamefont {Lampen-Kelley}, \citenamefont
  {Banerjee}, \citenamefont {Liu}, \citenamefont {Yan}, \citenamefont
  {Mandrus}, \citenamefont {Vojta},\ and\ \citenamefont {Nagler}}]{rucl3f5}%
  \BibitemOpen
  \bibfield  {author} {\bibinfo {author} {\bibfnamefont {C.}~\bibnamefont
  {Balz}}, \bibinfo {author} {\bibfnamefont {L.}~\bibnamefont {Janssen}},
  \bibinfo {author} {\bibfnamefont {P.}~\bibnamefont {Lampen-Kelley}}, \bibinfo
  {author} {\bibfnamefont {A.}~\bibnamefont {Banerjee}}, \bibinfo {author}
  {\bibfnamefont {Y.~H.}\ \bibnamefont {Liu}}, \bibinfo {author} {\bibfnamefont
  {J.-Q.}\ \bibnamefont {Yan}}, \bibinfo {author} {\bibfnamefont {D.~G.}\
  \bibnamefont {Mandrus}}, \bibinfo {author} {\bibfnamefont {M.}~\bibnamefont
  {Vojta}}, \ and\ \bibinfo {author} {\bibfnamefont {S.~E.}\ \bibnamefont
  {Nagler}},\ }\href {\doibase 10.1103/PhysRevB.103.174417} {\bibfield
  {journal} {\bibinfo  {journal} {Phys. Rev. B}\ }\textbf {\bibinfo {volume}
  {103}},\ \bibinfo {pages} {174417} (\bibinfo {year} {2021})}\BibitemShut
  {NoStop}%
\bibitem [{\citenamefont {Banerjee}\ \emph {et~al.}(2018)\citenamefont
  {Banerjee}, \citenamefont {Lampen-Kelley}, \citenamefont {Knolle},
  \citenamefont {Balz}, \citenamefont {Aczel}, \citenamefont {Winn},
  \citenamefont {Liu}, \citenamefont {Pajerowski}, \citenamefont {Yan},
  \citenamefont {Bridges}, \citenamefont {Savici}, \citenamefont {Chakoumakos},
  \citenamefont {Lumsden}, \citenamefont {Tennant}, \citenamefont {Moessner},
  \citenamefont {Mandrus},\ and\ \citenamefont {Nagler}}]{rucl3f6}%
  \BibitemOpen
  \bibfield  {author} {\bibinfo {author} {\bibfnamefont {A.}~\bibnamefont
  {Banerjee}}, \bibinfo {author} {\bibfnamefont {P.}~\bibnamefont
  {Lampen-Kelley}}, \bibinfo {author} {\bibfnamefont {J.}~\bibnamefont
  {Knolle}}, \bibinfo {author} {\bibfnamefont {C.}~\bibnamefont {Balz}},
  \bibinfo {author} {\bibfnamefont {A.~A.}\ \bibnamefont {Aczel}}, \bibinfo
  {author} {\bibfnamefont {B.}~\bibnamefont {Winn}}, \bibinfo {author}
  {\bibfnamefont {Y.}~\bibnamefont {Liu}}, \bibinfo {author} {\bibfnamefont
  {D.}~\bibnamefont {Pajerowski}}, \bibinfo {author} {\bibfnamefont
  {J.}~\bibnamefont {Yan}}, \bibinfo {author} {\bibfnamefont {C.~A.}\
  \bibnamefont {Bridges}}, \bibinfo {author} {\bibfnamefont {A.~T.}\
  \bibnamefont {Savici}}, \bibinfo {author} {\bibfnamefont {B.~C.}\
  \bibnamefont {Chakoumakos}}, \bibinfo {author} {\bibfnamefont {M.~D.}\
  \bibnamefont {Lumsden}}, \bibinfo {author} {\bibfnamefont {D.~A.}\
  \bibnamefont {Tennant}}, \bibinfo {author} {\bibfnamefont {R.}~\bibnamefont
  {Moessner}}, \bibinfo {author} {\bibfnamefont {D.~G.}\ \bibnamefont
  {Mandrus}}, \ and\ \bibinfo {author} {\bibfnamefont {S.~E.}\ \bibnamefont
  {Nagler}},\ }\href {\doibase 10.1038/s41535-018-0079-2} {\bibfield  {journal}
  {\bibinfo  {journal} {npj Quantum Materials}\ }\textbf {\bibinfo {volume}
  {3}},\ \bibinfo {pages} {8} (\bibinfo {year} {2018})}\BibitemShut {NoStop}%
\bibitem [{\citenamefont {Banerjee}\ \emph {et~al.}(2017)\citenamefont
  {Banerjee}, \citenamefont {Yan}, \citenamefont {Knolle}, \citenamefont
  {Bridges}, \citenamefont {Stone}, \citenamefont {Lumsden}, \citenamefont
  {Mandrus}, \citenamefont {Tennant}, \citenamefont {Moessner},\ and\
  \citenamefont {Nagler}}]{rclneutron}%
  \BibitemOpen
  \bibfield  {author} {\bibinfo {author} {\bibfnamefont {A.}~\bibnamefont
  {Banerjee}}, \bibinfo {author} {\bibfnamefont {J.}~\bibnamefont {Yan}},
  \bibinfo {author} {\bibfnamefont {J.}~\bibnamefont {Knolle}}, \bibinfo
  {author} {\bibfnamefont {C.~A.}\ \bibnamefont {Bridges}}, \bibinfo {author}
  {\bibfnamefont {M.~B.}\ \bibnamefont {Stone}}, \bibinfo {author}
  {\bibfnamefont {M.~D.}\ \bibnamefont {Lumsden}}, \bibinfo {author}
  {\bibfnamefont {D.~G.}\ \bibnamefont {Mandrus}}, \bibinfo {author}
  {\bibfnamefont {D.~A.}\ \bibnamefont {Tennant}}, \bibinfo {author}
  {\bibfnamefont {R.}~\bibnamefont {Moessner}}, \ and\ \bibinfo {author}
  {\bibfnamefont {S.~E.}\ \bibnamefont {Nagler}},\ }\href {\doibase
  10.1126/science.aah6015} {\bibfield  {journal} {\bibinfo  {journal}
  {Science}\ }\textbf {\bibinfo {volume} {356}},\ \bibinfo {pages} {1055}
  (\bibinfo {year} {2017})}\BibitemShut {NoStop}%
\bibitem [{\citenamefont {Yokoi}\ \emph {et~al.}(2021)\citenamefont {Yokoi},
  \citenamefont {Ma}, \citenamefont {Kasahara}, \citenamefont {Kasahara},
  \citenamefont {Shibauchi}, \citenamefont {Kurita}, \citenamefont {Tanaka},
  \citenamefont {Nasu}, \citenamefont {Motome}, \citenamefont {Hickey},
  \citenamefont {Trebst},\ and\ \citenamefont {Matsuda}}]{th1}%
  \BibitemOpen
  \bibfield  {author} {\bibinfo {author} {\bibfnamefont {T.}~\bibnamefont
  {Yokoi}}, \bibinfo {author} {\bibfnamefont {S.}~\bibnamefont {Ma}}, \bibinfo
  {author} {\bibfnamefont {Y.}~\bibnamefont {Kasahara}}, \bibinfo {author}
  {\bibfnamefont {S.}~\bibnamefont {Kasahara}}, \bibinfo {author}
  {\bibfnamefont {T.}~\bibnamefont {Shibauchi}}, \bibinfo {author}
  {\bibfnamefont {N.}~\bibnamefont {Kurita}}, \bibinfo {author} {\bibfnamefont
  {H.}~\bibnamefont {Tanaka}}, \bibinfo {author} {\bibfnamefont
  {J.}~\bibnamefont {Nasu}}, \bibinfo {author} {\bibfnamefont {Y.}~\bibnamefont
  {Motome}}, \bibinfo {author} {\bibfnamefont {C.}~\bibnamefont {Hickey}},
  \bibinfo {author} {\bibfnamefont {S.}~\bibnamefont {Trebst}}, \ and\ \bibinfo
  {author} {\bibfnamefont {Y.}~\bibnamefont {Matsuda}},\ }\href {\doibase
  10.1126/science.aay5551} {\bibfield  {journal} {\bibinfo  {journal}
  {Science}\ }\textbf {\bibinfo {volume} {373}},\ \bibinfo {pages} {568}
  (\bibinfo {year} {2021})}\BibitemShut {NoStop}%
\bibitem [{\citenamefont {Kasahara}\ \emph {et~al.}(2018)\citenamefont
  {Kasahara}, \citenamefont {Ohnishi}, \citenamefont {Mizukami}, \citenamefont
  {Tanaka}, \citenamefont {Ma}, \citenamefont {Sugii}, \citenamefont {Kurita},
  \citenamefont {Tanaka}, \citenamefont {Nasu}, \citenamefont {Motome},
  \citenamefont {Shibauchi},\ and\ \citenamefont {Matsuda}}]{th2}%
  \BibitemOpen
  \bibfield  {author} {\bibinfo {author} {\bibfnamefont {Y.}~\bibnamefont
  {Kasahara}}, \bibinfo {author} {\bibfnamefont {T.}~\bibnamefont {Ohnishi}},
  \bibinfo {author} {\bibfnamefont {Y.}~\bibnamefont {Mizukami}}, \bibinfo
  {author} {\bibfnamefont {O.}~\bibnamefont {Tanaka}}, \bibinfo {author}
  {\bibfnamefont {S.}~\bibnamefont {Ma}}, \bibinfo {author} {\bibfnamefont
  {K.}~\bibnamefont {Sugii}}, \bibinfo {author} {\bibfnamefont
  {N.}~\bibnamefont {Kurita}}, \bibinfo {author} {\bibfnamefont
  {H.}~\bibnamefont {Tanaka}}, \bibinfo {author} {\bibfnamefont
  {J.}~\bibnamefont {Nasu}}, \bibinfo {author} {\bibfnamefont {Y.}~\bibnamefont
  {Motome}}, \bibinfo {author} {\bibfnamefont {T.}~\bibnamefont {Shibauchi}}, \
  and\ \bibinfo {author} {\bibfnamefont {Y.}~\bibnamefont {Matsuda}},\ }\href
  {\doibase 10.1038/s41586-018-0274-0} {\bibfield  {journal} {\bibinfo
  {journal} {Nature}\ }\textbf {\bibinfo {volume} {559}},\ \bibinfo {pages}
  {227} (\bibinfo {year} {2018})}\BibitemShut {NoStop}%
\bibitem [{\citenamefont {Bruin}\ \emph
  {et~al.}(2022{\natexlab{a}})\citenamefont {Bruin}, \citenamefont {Claus},
  \citenamefont {Matsumoto}, \citenamefont {Kurita}, \citenamefont {Tanaka},\
  and\ \citenamefont {Takagi}}]{th3}%
  \BibitemOpen
  \bibfield  {author} {\bibinfo {author} {\bibfnamefont {J.~A.~N.}\
  \bibnamefont {Bruin}}, \bibinfo {author} {\bibfnamefont {R.~R.}\ \bibnamefont
  {Claus}}, \bibinfo {author} {\bibfnamefont {Y.}~\bibnamefont {Matsumoto}},
  \bibinfo {author} {\bibfnamefont {N.}~\bibnamefont {Kurita}}, \bibinfo
  {author} {\bibfnamefont {H.}~\bibnamefont {Tanaka}}, \ and\ \bibinfo {author}
  {\bibfnamefont {H.}~\bibnamefont {Takagi}},\ }\href {\doibase
  10.1038/s41567-021-01501-y} {\bibfield  {journal} {\bibinfo  {journal}
  {Nature Physics}\ }\textbf {\bibinfo {volume} {18}},\ \bibinfo {pages} {401}
  (\bibinfo {year} {2022}{\natexlab{a}})}\BibitemShut {NoStop}%
\bibitem [{\citenamefont {Czajka}\ \emph {et~al.}(2021)\citenamefont {Czajka},
  \citenamefont {Gao}, \citenamefont {Hirschberger}, \citenamefont
  {Lampen-Kelley}, \citenamefont {Banerjee}, \citenamefont {Yan}, \citenamefont
  {Mandrus}, \citenamefont {Nagler},\ and\ \citenamefont {Ong}}]{thos1}%
  \BibitemOpen
  \bibfield  {author} {\bibinfo {author} {\bibfnamefont {P.}~\bibnamefont
  {Czajka}}, \bibinfo {author} {\bibfnamefont {T.}~\bibnamefont {Gao}},
  \bibinfo {author} {\bibfnamefont {M.}~\bibnamefont {Hirschberger}}, \bibinfo
  {author} {\bibfnamefont {P.}~\bibnamefont {Lampen-Kelley}}, \bibinfo {author}
  {\bibfnamefont {A.}~\bibnamefont {Banerjee}}, \bibinfo {author}
  {\bibfnamefont {J.}~\bibnamefont {Yan}}, \bibinfo {author} {\bibfnamefont
  {D.~G.}\ \bibnamefont {Mandrus}}, \bibinfo {author} {\bibfnamefont {S.~E.}\
  \bibnamefont {Nagler}}, \ and\ \bibinfo {author} {\bibfnamefont {N.~P.}\
  \bibnamefont {Ong}},\ }\href {\doibase 10.1038/s41567-021-01243-x} {\bibfield
   {journal} {\bibinfo  {journal} {Nature Physics}\ }\textbf {\bibinfo {volume}
  {17}},\ \bibinfo {pages} {915} (\bibinfo {year} {2021})}\BibitemShut
  {NoStop}%
\bibitem [{\citenamefont {Bruin}\ \emph
  {et~al.}(2022{\natexlab{b}})\citenamefont {Bruin}, \citenamefont {Claus},
  \citenamefont {Matsumoto}, \citenamefont {Nuss}, \citenamefont {Laha},
  \citenamefont {Lotsch}, \citenamefont {Kurita}, \citenamefont {Tanaka},\ and\
  \citenamefont {Takagi}}]{thos2}%
  \BibitemOpen
  \bibfield  {author} {\bibinfo {author} {\bibfnamefont {J.~A.~N.}\
  \bibnamefont {Bruin}}, \bibinfo {author} {\bibfnamefont {R.~R.}\ \bibnamefont
  {Claus}}, \bibinfo {author} {\bibfnamefont {Y.}~\bibnamefont {Matsumoto}},
  \bibinfo {author} {\bibfnamefont {J.}~\bibnamefont {Nuss}}, \bibinfo {author}
  {\bibfnamefont {S.}~\bibnamefont {Laha}}, \bibinfo {author} {\bibfnamefont
  {B.~V.}\ \bibnamefont {Lotsch}}, \bibinfo {author} {\bibfnamefont
  {N.}~\bibnamefont {Kurita}}, \bibinfo {author} {\bibfnamefont
  {H.}~\bibnamefont {Tanaka}}, \ and\ \bibinfo {author} {\bibfnamefont
  {H.}~\bibnamefont {Takagi}},\ }\href {\doibase 10.1063/5.0101377} {\bibfield
  {journal} {\bibinfo  {journal} {APL Materials}\ }\textbf {\bibinfo {volume}
  {10}} (\bibinfo {year} {2022}{\natexlab{b}}),\ 10.1063/5.0101377},\ \bibinfo
  {note} {090703}\BibitemShut {NoStop}%
\bibitem [{\citenamefont {Czajka}\ \emph {et~al.}(2023)\citenamefont {Czajka},
  \citenamefont {Gao}, \citenamefont {Hirschberger}, \citenamefont
  {Lampen-Kelley}, \citenamefont {Banerjee}, \citenamefont {Quirk},
  \citenamefont {Mandrus}, \citenamefont {Nagler},\ and\ \citenamefont
  {Ong}}]{th4}%
  \BibitemOpen
  \bibfield  {author} {\bibinfo {author} {\bibfnamefont {P.}~\bibnamefont
  {Czajka}}, \bibinfo {author} {\bibfnamefont {T.}~\bibnamefont {Gao}},
  \bibinfo {author} {\bibfnamefont {M.}~\bibnamefont {Hirschberger}}, \bibinfo
  {author} {\bibfnamefont {P.}~\bibnamefont {Lampen-Kelley}}, \bibinfo {author}
  {\bibfnamefont {A.}~\bibnamefont {Banerjee}}, \bibinfo {author}
  {\bibfnamefont {N.}~\bibnamefont {Quirk}}, \bibinfo {author} {\bibfnamefont
  {D.~G.}\ \bibnamefont {Mandrus}}, \bibinfo {author} {\bibfnamefont {S.~E.}\
  \bibnamefont {Nagler}}, \ and\ \bibinfo {author} {\bibfnamefont {N.~P.}\
  \bibnamefont {Ong}},\ }\href {\doibase 10.1038/s41563-022-01397-w} {\bibfield
   {journal} {\bibinfo  {journal} {Nature Materials}\ }\textbf {\bibinfo
  {volume} {22}},\ \bibinfo {pages} {36} (\bibinfo {year} {2023})}\BibitemShut
  {NoStop}%
\bibitem [{\citenamefont {Lefran\ifmmode~\mbox{\c{c}}\else \c{c}\fi{}ois}\
  \emph {et~al.}(2023)\citenamefont {Lefran\ifmmode~\mbox{\c{c}}\else
  \c{c}\fi{}ois}, \citenamefont {Baglo}, \citenamefont {Barth\'elemy},
  \citenamefont {Kim}, \citenamefont {Kim},\ and\ \citenamefont
  {Taillefer}}]{thos3}%
  \BibitemOpen
  \bibfield  {author} {\bibinfo {author} {\bibfnamefont {E.}~\bibnamefont
  {Lefran\ifmmode~\mbox{\c{c}}\else \c{c}\fi{}ois}}, \bibinfo {author}
  {\bibfnamefont {J.}~\bibnamefont {Baglo}}, \bibinfo {author} {\bibfnamefont
  {Q.}~\bibnamefont {Barth\'elemy}}, \bibinfo {author} {\bibfnamefont
  {S.}~\bibnamefont {Kim}}, \bibinfo {author} {\bibfnamefont {Y.-J.}\
  \bibnamefont {Kim}}, \ and\ \bibinfo {author} {\bibfnamefont
  {L.}~\bibnamefont {Taillefer}},\ }\href {\doibase
  10.1103/PhysRevB.107.064408} {\bibfield  {journal} {\bibinfo  {journal}
  {Phys. Rev. B}\ }\textbf {\bibinfo {volume} {107}},\ \bibinfo {pages}
  {064408} (\bibinfo {year} {2023})}\BibitemShut {NoStop}%
\bibitem [{\citenamefont {Tanaka}\ \emph {et~al.}(2022)\citenamefont {Tanaka},
  \citenamefont {Mizukami}, \citenamefont {Harasawa}, \citenamefont
  {Hashimoto}, \citenamefont {Hwang}, \citenamefont {Kurita}, \citenamefont
  {Tanaka}, \citenamefont {Fujimoto}, \citenamefont {Matsuda}, \citenamefont
  {Moon},\ and\ \citenamefont {Shibauchi}}]{rucl3cp}%
  \BibitemOpen
  \bibfield  {author} {\bibinfo {author} {\bibfnamefont {O.}~\bibnamefont
  {Tanaka}}, \bibinfo {author} {\bibfnamefont {Y.}~\bibnamefont {Mizukami}},
  \bibinfo {author} {\bibfnamefont {R.}~\bibnamefont {Harasawa}}, \bibinfo
  {author} {\bibfnamefont {K.}~\bibnamefont {Hashimoto}}, \bibinfo {author}
  {\bibfnamefont {K.}~\bibnamefont {Hwang}}, \bibinfo {author} {\bibfnamefont
  {N.}~\bibnamefont {Kurita}}, \bibinfo {author} {\bibfnamefont
  {H.}~\bibnamefont {Tanaka}}, \bibinfo {author} {\bibfnamefont
  {S.}~\bibnamefont {Fujimoto}}, \bibinfo {author} {\bibfnamefont
  {Y.}~\bibnamefont {Matsuda}}, \bibinfo {author} {\bibfnamefont {E.-G.}\
  \bibnamefont {Moon}}, \ and\ \bibinfo {author} {\bibfnamefont
  {T.}~\bibnamefont {Shibauchi}},\ }\href {\doibase 10.1038/s41567-021-01488-6}
  {\bibfield  {journal} {\bibinfo  {journal} {Nature Physics}\ }\textbf
  {\bibinfo {volume} {18}},\ \bibinfo {pages} {429} (\bibinfo {year}
  {2022})}\BibitemShut {NoStop}%
\bibitem [{\citenamefont {Zhang}()}]{supp}%
  \BibitemOpen
  \bibfield  {author} {\bibinfo {author} {\bibfnamefont {H.}~\bibnamefont
  {Zhang}},\ }\href@noop {} {\bibinfo  {journal} {Supplemental material to
  "Anisotropy ..."}\ }\BibitemShut {NoStop}%
\bibitem [{\citenamefont {Imamura}\ \emph {et~al.}(2023)\citenamefont
  {Imamura}, \citenamefont {Suetsugu}, \citenamefont {Mizukami}, \citenamefont
  {Yoshida}, \citenamefont {Hashimoto}, \citenamefont {Ohtsuka}, \citenamefont
  {Kasahara}, \citenamefont {Kurita}, \citenamefont {Tanaka}, \citenamefont
  {Noh}, \citenamefont {Nasu}, \citenamefont {Moon}, \citenamefont {Matsuda},\
  and\ \citenamefont {Shibauchi}}]{ang3}%
  \BibitemOpen
\bibfield  {journal} {  }\bibfield  {author} {\bibinfo {author} {\bibfnamefont
  {K.}~\bibnamefont {Imamura}}, \bibinfo {author} {\bibfnamefont
  {S.}~\bibnamefont {Suetsugu}}, \bibinfo {author} {\bibfnamefont
  {Y.}~\bibnamefont {Mizukami}}, \bibinfo {author} {\bibfnamefont
  {Y.}~\bibnamefont {Yoshida}}, \bibinfo {author} {\bibfnamefont
  {K.}~\bibnamefont {Hashimoto}}, \bibinfo {author} {\bibfnamefont
  {K.}~\bibnamefont {Ohtsuka}}, \bibinfo {author} {\bibfnamefont
  {Y.}~\bibnamefont {Kasahara}}, \bibinfo {author} {\bibfnamefont
  {N.}~\bibnamefont {Kurita}}, \bibinfo {author} {\bibfnamefont
  {H.}~\bibnamefont {Tanaka}}, \bibinfo {author} {\bibfnamefont
  {P.}~\bibnamefont {Noh}}, \bibinfo {author} {\bibfnamefont {J.}~\bibnamefont
  {Nasu}}, \bibinfo {author} {\bibfnamefont {E.~G.}\ \bibnamefont {Moon}},
  \bibinfo {author} {\bibfnamefont {Y.}~\bibnamefont {Matsuda}}, \ and\
  \bibinfo {author} {\bibfnamefont {T.}~\bibnamefont {Shibauchi}},\ }\href@noop
  {} {\enquote {\bibinfo {title} {Majorana-fermion origin of the planar thermal
  hall effect in the kitaev magnet $\alpha$-rucl$_3$},}\ } (\bibinfo {year}
  {2023}),\ \Eprint {http://arxiv.org/abs/2305.10619} {arXiv:2305.10619
  [cond-mat.str-el]} \BibitemShut {NoStop}%
\bibitem [{\citenamefont {Mu}\ \emph {et~al.}(2022)\citenamefont {Mu},
  \citenamefont {Dixit}, \citenamefont {Wang}, \citenamefont {Abernathy},
  \citenamefont {Cao}, \citenamefont {Nagler}, \citenamefont {Yan},
  \citenamefont {Lampen-Kelley}, \citenamefont {Mandrus}, \citenamefont
  {Polanco}, \citenamefont {Liang}, \citenamefont {Hal\'asz}, \citenamefont
  {Cheng}, \citenamefont {Banerjee},\ and\ \citenamefont {Berlijn}}]{phone}%
  \BibitemOpen
  \bibfield  {author} {\bibinfo {author} {\bibfnamefont {S.}~\bibnamefont
  {Mu}}, \bibinfo {author} {\bibfnamefont {K.~D.}\ \bibnamefont {Dixit}},
  \bibinfo {author} {\bibfnamefont {X.}~\bibnamefont {Wang}}, \bibinfo {author}
  {\bibfnamefont {D.~L.}\ \bibnamefont {Abernathy}}, \bibinfo {author}
  {\bibfnamefont {H.}~\bibnamefont {Cao}}, \bibinfo {author} {\bibfnamefont
  {S.~E.}\ \bibnamefont {Nagler}}, \bibinfo {author} {\bibfnamefont
  {J.}~\bibnamefont {Yan}}, \bibinfo {author} {\bibfnamefont {P.}~\bibnamefont
  {Lampen-Kelley}}, \bibinfo {author} {\bibfnamefont {D.}~\bibnamefont
  {Mandrus}}, \bibinfo {author} {\bibfnamefont {C.~A.}\ \bibnamefont
  {Polanco}}, \bibinfo {author} {\bibfnamefont {L.}~\bibnamefont {Liang}},
  \bibinfo {author} {\bibfnamefont {G.~B.}\ \bibnamefont {Hal\'asz}}, \bibinfo
  {author} {\bibfnamefont {Y.}~\bibnamefont {Cheng}}, \bibinfo {author}
  {\bibfnamefont {A.}~\bibnamefont {Banerjee}}, \ and\ \bibinfo {author}
  {\bibfnamefont {T.}~\bibnamefont {Berlijn}},\ }\href {\doibase
  10.1103/PhysRevResearch.4.013067} {\bibfield  {journal} {\bibinfo  {journal}
  {Phys. Rev. Res.}\ }\textbf {\bibinfo {volume} {4}},\ \bibinfo {pages}
  {013067} (\bibinfo {year} {2022})}\BibitemShut {NoStop}%
\bibitem [{\citenamefont {Ponomaryov}\ \emph {et~al.}(2017)\citenamefont
  {Ponomaryov}, \citenamefont {Schulze}, \citenamefont {Wosnitza},
  \citenamefont {Lampen-Kelley}, \citenamefont {Banerjee}, \citenamefont {Yan},
  \citenamefont {Bridges}, \citenamefont {Mandrus}, \citenamefont {Nagler},
  \citenamefont {Kolezhuk},\ and\ \citenamefont {Zvyagin}}]{meg1}%
  \BibitemOpen
  \bibfield  {author} {\bibinfo {author} {\bibfnamefont {A.~N.}\ \bibnamefont
  {Ponomaryov}}, \bibinfo {author} {\bibfnamefont {E.}~\bibnamefont {Schulze}},
  \bibinfo {author} {\bibfnamefont {J.}~\bibnamefont {Wosnitza}}, \bibinfo
  {author} {\bibfnamefont {P.}~\bibnamefont {Lampen-Kelley}}, \bibinfo {author}
  {\bibfnamefont {A.}~\bibnamefont {Banerjee}}, \bibinfo {author}
  {\bibfnamefont {J.-Q.}\ \bibnamefont {Yan}}, \bibinfo {author} {\bibfnamefont
  {C.~A.}\ \bibnamefont {Bridges}}, \bibinfo {author} {\bibfnamefont {D.~G.}\
  \bibnamefont {Mandrus}}, \bibinfo {author} {\bibfnamefont {S.~E.}\
  \bibnamefont {Nagler}}, \bibinfo {author} {\bibfnamefont {A.~K.}\
  \bibnamefont {Kolezhuk}}, \ and\ \bibinfo {author} {\bibfnamefont {S.~A.}\
  \bibnamefont {Zvyagin}},\ }\href {\doibase 10.1103/PhysRevB.96.241107}
  {\bibfield  {journal} {\bibinfo  {journal} {Phys. Rev. B}\ }\textbf {\bibinfo
  {volume} {96}},\ \bibinfo {pages} {241107} (\bibinfo {year}
  {2017})}\BibitemShut {NoStop}%
\bibitem [{\citenamefont {Ponomaryov}\ \emph {et~al.}(2020)\citenamefont
  {Ponomaryov}, \citenamefont {Zviagina}, \citenamefont {Wosnitza},
  \citenamefont {Lampen-Kelley}, \citenamefont {Banerjee}, \citenamefont {Yan},
  \citenamefont {Bridges}, \citenamefont {Mandrus}, \citenamefont {Nagler},\
  and\ \citenamefont {Zvyagin}}]{meg2}%
  \BibitemOpen
  \bibfield  {author} {\bibinfo {author} {\bibfnamefont {A.~N.}\ \bibnamefont
  {Ponomaryov}}, \bibinfo {author} {\bibfnamefont {L.}~\bibnamefont
  {Zviagina}}, \bibinfo {author} {\bibfnamefont {J.}~\bibnamefont {Wosnitza}},
  \bibinfo {author} {\bibfnamefont {P.}~\bibnamefont {Lampen-Kelley}}, \bibinfo
  {author} {\bibfnamefont {A.}~\bibnamefont {Banerjee}}, \bibinfo {author}
  {\bibfnamefont {J.-Q.}\ \bibnamefont {Yan}}, \bibinfo {author} {\bibfnamefont
  {C.~A.}\ \bibnamefont {Bridges}}, \bibinfo {author} {\bibfnamefont {D.~G.}\
  \bibnamefont {Mandrus}}, \bibinfo {author} {\bibfnamefont {S.~E.}\
  \bibnamefont {Nagler}}, \ and\ \bibinfo {author} {\bibfnamefont {S.~A.}\
  \bibnamefont {Zvyagin}},\ }\href {\doibase 10.1103/PhysRevLett.125.037202}
  {\bibfield  {journal} {\bibinfo  {journal} {Phys. Rev. Lett.}\ }\textbf
  {\bibinfo {volume} {125}},\ \bibinfo {pages} {037202} (\bibinfo {year}
  {2020})}\BibitemShut {NoStop}%
\bibitem [{\citenamefont {Winter}\ \emph {et~al.}(2018)\citenamefont {Winter},
  \citenamefont {Riedl}, \citenamefont {Kaib}, \citenamefont {Coldea},\ and\
  \citenamefont {Valent\'{\i}}}]{conti1}%
  \BibitemOpen
  \bibfield  {author} {\bibinfo {author} {\bibfnamefont {S.~M.}\ \bibnamefont
  {Winter}}, \bibinfo {author} {\bibfnamefont {K.}~\bibnamefont {Riedl}},
  \bibinfo {author} {\bibfnamefont {D.}~\bibnamefont {Kaib}}, \bibinfo {author}
  {\bibfnamefont {R.}~\bibnamefont {Coldea}}, \ and\ \bibinfo {author}
  {\bibfnamefont {R.}~\bibnamefont {Valent\'{\i}}},\ }\href {\doibase
  10.1103/PhysRevLett.120.077203} {\bibfield  {journal} {\bibinfo  {journal}
  {Phys. Rev. Lett.}\ }\textbf {\bibinfo {volume} {120}},\ \bibinfo {pages}
  {077203} (\bibinfo {year} {2018})}\BibitemShut {NoStop}%
\bibitem [{\citenamefont {Wellm}\ \emph {et~al.}(2018)\citenamefont {Wellm},
  \citenamefont {Zeisner}, \citenamefont {Alfonsov}, \citenamefont {Wolter},
  \citenamefont {Roslova}, \citenamefont {Isaeva}, \citenamefont {Doert},
  \citenamefont {Vojta}, \citenamefont {B\"uchner},\ and\ \citenamefont
  {Kataev}}]{conti2}%
  \BibitemOpen
  \bibfield  {author} {\bibinfo {author} {\bibfnamefont {C.}~\bibnamefont
  {Wellm}}, \bibinfo {author} {\bibfnamefont {J.}~\bibnamefont {Zeisner}},
  \bibinfo {author} {\bibfnamefont {A.}~\bibnamefont {Alfonsov}}, \bibinfo
  {author} {\bibfnamefont {A.~U.~B.}\ \bibnamefont {Wolter}}, \bibinfo {author}
  {\bibfnamefont {M.}~\bibnamefont {Roslova}}, \bibinfo {author} {\bibfnamefont
  {A.}~\bibnamefont {Isaeva}}, \bibinfo {author} {\bibfnamefont
  {T.}~\bibnamefont {Doert}}, \bibinfo {author} {\bibfnamefont
  {M.}~\bibnamefont {Vojta}}, \bibinfo {author} {\bibfnamefont
  {B.}~\bibnamefont {B\"uchner}}, \ and\ \bibinfo {author} {\bibfnamefont
  {V.}~\bibnamefont {Kataev}},\ }\href {\doibase 10.1103/PhysRevB.98.184408}
  {\bibfield  {journal} {\bibinfo  {journal} {Phys. Rev. B}\ }\textbf {\bibinfo
  {volume} {98}},\ \bibinfo {pages} {184408} (\bibinfo {year}
  {2018})}\BibitemShut {NoStop}%
\bibitem [{\citenamefont {Wan}\ \emph {et~al.}(2023)\citenamefont {Wan},
  \citenamefont {Zheliuk}, \citenamefont {Yuan}, \citenamefont {Peng},
  \citenamefont {Zhang}, \citenamefont {Liang}, \citenamefont {Zeitler},
  \citenamefont {Wiedmann}, \citenamefont {Hussey}, \citenamefont {Palstra},\
  and\ \citenamefont {Ye}}]{fflo}%
  \BibitemOpen
  \bibfield  {author} {\bibinfo {author} {\bibfnamefont {P.}~\bibnamefont
  {Wan}}, \bibinfo {author} {\bibfnamefont {O.}~\bibnamefont {Zheliuk}},
  \bibinfo {author} {\bibfnamefont {N.~F.~Q.}\ \bibnamefont {Yuan}}, \bibinfo
  {author} {\bibfnamefont {X.}~\bibnamefont {Peng}}, \bibinfo {author}
  {\bibfnamefont {L.}~\bibnamefont {Zhang}}, \bibinfo {author} {\bibfnamefont
  {M.}~\bibnamefont {Liang}}, \bibinfo {author} {\bibfnamefont
  {U.}~\bibnamefont {Zeitler}}, \bibinfo {author} {\bibfnamefont
  {S.}~\bibnamefont {Wiedmann}}, \bibinfo {author} {\bibfnamefont {N.~E.}\
  \bibnamefont {Hussey}}, \bibinfo {author} {\bibfnamefont {T.~T.~M.}\
  \bibnamefont {Palstra}}, \ and\ \bibinfo {author} {\bibfnamefont
  {J.}~\bibnamefont {Ye}},\ }\href {\doibase 10.1038/s41586-023-05967-z}
  {\bibfield  {journal} {\bibinfo  {journal} {Nature}\ }\textbf {\bibinfo
  {volume} {619}},\ \bibinfo {pages} {46} (\bibinfo {year} {2023})}\BibitemShut
  {NoStop}%
\bibitem [{\citenamefont {Li}\ \emph {et~al.}(2023)\citenamefont {Li},
  \citenamefont {Yi}, \citenamefont {Wang}, \citenamefont {Yu}, \citenamefont
  {Shen}, \citenamefont {Yan},\ and\ \citenamefont {Wang}}]{kvs}%
  \BibitemOpen
  \bibfield  {author} {\bibinfo {author} {\bibfnamefont {L.}~\bibnamefont
  {Li}}, \bibinfo {author} {\bibfnamefont {E.}~\bibnamefont {Yi}}, \bibinfo
  {author} {\bibfnamefont {B.}~\bibnamefont {Wang}}, \bibinfo {author}
  {\bibfnamefont {G.}~\bibnamefont {Yu}}, \bibinfo {author} {\bibfnamefont
  {B.}~\bibnamefont {Shen}}, \bibinfo {author} {\bibfnamefont {Z.}~\bibnamefont
  {Yan}}, \ and\ \bibinfo {author} {\bibfnamefont {M.}~\bibnamefont {Wang}},\
  }\href {\doibase 10.1038/s41535-022-00534-7} {\bibfield  {journal} {\bibinfo
  {journal} {npj Quantum Materials}\ }\textbf {\bibinfo {volume} {8}},\
  \bibinfo {pages} {2} (\bibinfo {year} {2023})}\BibitemShut {NoStop}%
\bibitem [{\citenamefont {Seitz}(1950)}]{mr}%
  \BibitemOpen
  \bibfield  {author} {\bibinfo {author} {\bibfnamefont {F.}~\bibnamefont
  {Seitz}},\ }\href {\doibase 10.1103/PhysRev.79.372} {\bibfield  {journal}
  {\bibinfo  {journal} {Phys. Rev.}\ }\textbf {\bibinfo {volume} {79}},\
  \bibinfo {pages} {372} (\bibinfo {year} {1950})}\BibitemShut {NoStop}%
\bibitem [{\citenamefont {Yang}\ \emph {et~al.}(2020)\citenamefont {Yang},
  \citenamefont {Chang},\ and\ \citenamefont {Parkin}}]{Bifilm}%
  \BibitemOpen
  \bibfield  {author} {\bibinfo {author} {\bibfnamefont {S.-Y.}\ \bibnamefont
  {Yang}}, \bibinfo {author} {\bibfnamefont {K.}~\bibnamefont {Chang}}, \ and\
  \bibinfo {author} {\bibfnamefont {S.~S.~P.}\ \bibnamefont {Parkin}},\ }\href
  {\doibase 10.1103/PhysRevResearch.2.022029} {\bibfield  {journal} {\bibinfo
  {journal} {Phys. Rev. Res.}\ }\textbf {\bibinfo {volume} {2}},\ \bibinfo
  {pages} {022029} (\bibinfo {year} {2020})}\BibitemShut {NoStop}%
\bibitem [{\citenamefont {Rout}\ \emph {et~al.}(2017)\citenamefont {Rout},
  \citenamefont {Agireen}, \citenamefont {Maniv}, \citenamefont {Goldstein},\
  and\ \citenamefont {Dagan}}]{laosto}%
  \BibitemOpen
  \bibfield  {author} {\bibinfo {author} {\bibfnamefont {P.~K.}\ \bibnamefont
  {Rout}}, \bibinfo {author} {\bibfnamefont {I.}~\bibnamefont {Agireen}},
  \bibinfo {author} {\bibfnamefont {E.}~\bibnamefont {Maniv}}, \bibinfo
  {author} {\bibfnamefont {M.}~\bibnamefont {Goldstein}}, \ and\ \bibinfo
  {author} {\bibfnamefont {Y.}~\bibnamefont {Dagan}},\ }\href {\doibase
  10.1103/PhysRevB.95.241107} {\bibfield  {journal} {\bibinfo  {journal} {Phys.
  Rev. B}\ }\textbf {\bibinfo {volume} {95}},\ \bibinfo {pages} {241107}
  (\bibinfo {year} {2017})}\BibitemShut {NoStop}%
\end{thebibliography}

\end{document}